\documentclass[aps,prl,tightenlines,onecolumn,superscriptaddress,showpacs,report]{revtex4}
\usepackage{amssymb}
\usepackage{color,graphicx}
\usepackage{graphicx}
\usepackage{lmodern}
\usepackage{amsmath}
\usepackage{amsbsy}
\usepackage{amsthm}
\usepackage{float}
\usepackage{bbm}
\usepackage{bm}

\begin{document}
\title{Investigation of Stimulated Brillouin Scattering in Laser-Plasma Interactions}

\author{Mehdi Ghalamkarian Nejad}
\email[]{m.ghalamkarian@ph.iut.ac.ir}
\affiliation{Department of Physics, Isfahan University of Technology,Isfahan 8415683111, Iran}
\author{Milad Ghadimi}
\email[]{milad.ghadimi@ph.iut.ac.ir}
\affiliation{Department of Physics, Isfahan University of Technology,Isfahan 8415683111, Iran}
\begin{abstract}
In this paper we present our numerical simulation results on the Stimulated Brillouin Scattering (SBS) with injection of an ordinary mode (O-mode) electromagnetic wave (our pump wave) with frequencies $70$ GHz and $110$ GHz. Solving the Fourier transformed Vlasov equation in the velocity space, creates a profile for distribution function. Time evolution of the distribution function is investigated as well. Considering an average density for plasma fusion $(n_{0} \approx 10^{19} m^{-3})$, we gain a profile for density. Then two-dimensional instability rate for SBS is obtained. So, the fluctuation of distribution function affects density and again density affects instability rate. Increasing the incident light wave frequency causes the instability growth rate to decrease. Time evolution shows a clear damping for instability rate, since the pump wave's energy is absorbed in plasma (plasma heating). Furthermore, changing Landau damping for ion acoustic waves (IAW) by changing ion-to-electron temperature ratio is presented as well, because this damping is more dominant at high temperatures.
\end{abstract}

\pacs{valid number}
\maketitle
\section{I. Introduction}
There are some methods for plasma heating and the best way in very high temperatures is using the radio frequency (RF) wave heating. A high-power laser wave serving as a pump, can induce some parametric instabilities in plasma which one of them is SBS ~\cite{Elil}. Propagating the incident light wave through a plasma whose density is rippled along the direction of propagation by density fluctuations associated with the ion-acoustic wave (IAW), creates a small amplitude scattered electromagnetic (EM) wave, with all waves obeying Manley-Rowe relations (conservation laws for energy and momentum). Superposition of the incident pump EM wave and the scattered EM wave (wave- wave interaction) leads to the creation of Ponderomotive force. This force creates parametric instabilities. 
Brillouin instability is the coupling of a large amplitude light wave (with the frequency $\omega_0$ and wave number $k_0$) into a scattered light wave (with the frequency $\omega_{s}$ and wave number $k_{s}$) plus an ion acoustic wave (with the frequency $\omega_{iaw}$ and wave number $k_{iaw}$). Manley-Rowe relations leads to:
\begin{equation}
\omega_{0}=\omega_{s}+\omega_{iaw}
\end{equation}
\begin{equation}
k_{0}=k_{s}+k_{iaw}
\end{equation}
Since the ion acoustic frequency is less than the light wave frequency $(\omega_{iaw}\ll \omega_{0})$, this instability occurs in low-density plasma,$(n\ll n_{c}=\frac{(\varepsilon_{0}m_{e})}{e^2}\omega^2)$. Furthermore, almost all energy transfers to the scattered wave. $(Eenrgy \propto frequency)$. Therefore, investigation of this instability is very important, since it can decrease the absorption of energy into the plasma ~\cite{Kru}. \\
In the weak field limit $(\omega=\omega_{iaw}+i\gamma$ , with $\gamma \ll \omega_{iaw })$ and $\omega_{iaw}=k_{iaw}c_{s}$, where  $c_{s}$ is ion-acoustic velocity, the dispersion relation is:
\begin{equation}
\omega^2-k^2 c_{s}^2=\frac{\kappa^2 v_{os}^2}{4} \omega_{pi}^2 [\frac{1}{D(k-k_{0}.\omega - \omega_{0})}+\frac{1}{D(k+k_{0}.\omega + \omega_{0})}]
\end{equation}
which
\begin{equation}
D(k,\omega)=\omega^2 - k^2 -\omega_{p\epsilon}^2
\end{equation}
 \textquotedblleft Maximum growth rate occurs when the ion acoustic wave (IAW) and downshifted light wave are both resonant"~\cite{Rub} leads to $k_{iaw}$=$2k_{0} (cos\alpha-\frac{c_{s}}{\sqrt{{1-{n_{e}}/{n_{c}}}}} \frac{1}{c})$ and the growth rate is:
\begin{equation}
\gamma ^2=\frac{k_{0} v_{os} \omega_{pi}^2}{8 \omega_{0} c_{s}} cos\beta
\end{equation}
Here $\alpha$ is the angle between incident light wave and ion-acoustic wave. For maximum growth rate (backscattering) $\alpha$, the angle between incident light wave and scattered light wave is equal $\pi$. Since $k_{s}=k_{0}$, $2\beta+\alpha=\pi$. Therefore:
\begin{equation}
\gamma_{max}=\frac{1}{2\sqrt{2}} \frac{k_{0} v_{os} \omega_{pi}}{\sqrt{\omega_{0}k_{0}c_{s}}}
\end{equation}
Where $ v_{os}=\frac{e}{m_{e}\omega_{0}} \sqrt{\frac{8I}{c}}$, is oscillatory speed and I is laser intensity and $\omega_{pi}$ is ion plasma frequency ~\cite{Ga}.

\section{II.Evolution of Distribution Function:}
 \textquotedblleft A problem with the Vlasov equation is its tendency of becoming oscillatory in velocity space, potentially giving rise to recurrence effects where parts of the initial condition artificially reappear in the numerical solution"~\cite{Elia}. So, for numerical simulation we solve Fourier transformed Vlasov equation in the velocity space and time steps are gained by fourth order Runge-Kutta in FORTRAN program. For simplicity, the program produces two-dimensional distribution function $f(x,y,v_x,v_y,t)$. Letting $(n_{0} \approx 10^{19} m^{-3})$ (average density for plasma fusion like NSTX,MAST), we can gain a profile for density. By this new profile, electron and ion frequency, Debye length, instability growth rate profiles have been gained as well.\\
The basic equation for description of collisionless plasma is Vlasov equation which its non-relativistic form is written:
\begin{equation}
\frac{\partial f_{\alpha}}{\partial t}+v.\nabla_{r}f_{\alpha}+\frac{F_{\alpha}}{m_{\alpha}}.\nabla_{v}f_{\alpha}=0
\end{equation}
By using the Fourier transform pair, the Fourier transformed Vlasov equation is:
\begin{equation}
\frac{\partial \hat{f_{\alpha}}}{\partial t}-i\nabla_{x}.\nabla_{\eta} \hat{f_{\alpha}}-\frac{q_{\alpha}}{m_{\alpha}}\big(iE.\eta \hat{f_{\alpha}}+\nabla_{\eta}.\{[(B+B_{ext})\times \eta]\hat{f_{\alpha}}\})
\end{equation}
In which the velocity variable $v$ is transferred into a new variable $\eta$ and the distribution function $f(r,v,t)$ is changed to a new, complex valued, function $f(r,\eta,t)$.\\
We use periodic condition in space and discretize the problem on a rectangular, equidistant grid. The known variables in $r$ space are discretized as $x_{i_{1}}$=$i_{1} \Delta x $, $i_{1}$=$0$,$1$,\textellipsis,$N_{x-1}$ and $y_{i_{2}}$=$i_{2} \Delta y$, $i_{1}$=$0$,$1$,\textellipsis,$N_{y-1}$ the grid sizes are $\Delta x=\frac{L_{1}}{N_{y}}$, $\Delta y$=$\frac{L_{2}}{N_{y}}$where $L_1$ and $L_2$ are the domain sizes in the $x$ and $y$ directions, respectively.We use the domain size $0 \leq \eta_{1} \leq \eta_{1,\alpha,max}$and $-\eta_{2,\alpha,max} \leq  \eta_{2} \leq \eta_{2,\alpha,max}$for particle species $\alpha$(ions or electrons). The Fourier transformed velocity variables are discretized as:
\begin{equation}
\eta_{1,\alpha,j_{1}}=j_{1} \Delta \eta_{1,\alpha}     \    ,   \  \   j_{1}=0,1,\textellipsis,N_{\eta_{1}}
\end{equation}
\begin{equation}
\eta_{2,\alpha,j_{2}}=j_{2} \Delta \eta_{2,\alpha}    \     ,   \   \   j_{2}=0,1,\textellipsis,N_{\eta_{1}}
\end{equation}
Where the grid sizes now are:
\begin{equation}
\Delta \eta_{1,\alpha}  =\frac{\eta_{\alpha 1,max}}{N_{\eta_{1}}}
\end{equation}
\begin{equation}
\Delta \eta_{2,\alpha}  =\frac{\eta_{\alpha 2,max}}{N_{\eta_{2}}}
\end{equation}
And time discretizes as  $ t_k$=$t_{k-1}$+$\Delta t_k$   ,   $t_0=0$
, where $k=1 ,2 ,\textellipsis,N_t $   and $\Delta t=\frac{t_{end}}{N_t}$\\
Furthermore, we used dimensionless quantities (primed) for numerical simulation.\\    
\begin{equation}
t=\omega_{pe}^{-1}t'
\end{equation}
\begin{equation}
v_x=v_{th,e} v'_x     ,     v_y=v_{th,e} v'_y      
\end{equation}
\begin{equation}
x=\lambda_D x'      ,    y=\lambda_D y'
\end{equation}
Where $v_{th,e}=\sqrt{\frac{k_B T_e}{m_e}}$ is electron thermal velocity and $\lambda_D=\sqrt{\frac{v_{th}}{\omega_{pe}}}$ is Debye length.~\cite{Elia,Elias}

\section{III. Simulation Method:}
The plasma is unmagnetized since \textquotedblleft For spherical tokamaks with low B (magnetic field), wave injection is applied but for tokamaks with high B this issue is not necessary because cyclotron frequency is high and itself leads to plasma heating." ~\cite{Nagh}\\
The number of cells in x-direction is considered  $x/\lambda_D =50$ , for more simplicity, $y/\lambda_D =1$ in y-direction, $N_t=2000$ for time steps, $\eta_{1,max}=30$ and $\eta_{2,max}=60$. Since strong ion Landau damping occurs when $T_i$ approaches $T_e$ and we are in non-relativistic limit, following initial conditions were applied as well:
\begin{equation}
K_{B}T_{e}=K_{B}T_{i}=1 Kev
\end{equation}
\begin{equation}
I_{1}=10^{17} \frac{W}{cm^2}
\end{equation}
Dependence of instability growth rate on $\alpha$ (the angle between incident light wave and scattered light wave) is investigated and it corresponds with our expectations. So, all investigations have been done just for this situation (backscattering) which is a big concern for heating.
\begin{figure}[H]
\centering
\includegraphics[width=0.75\textwidth]{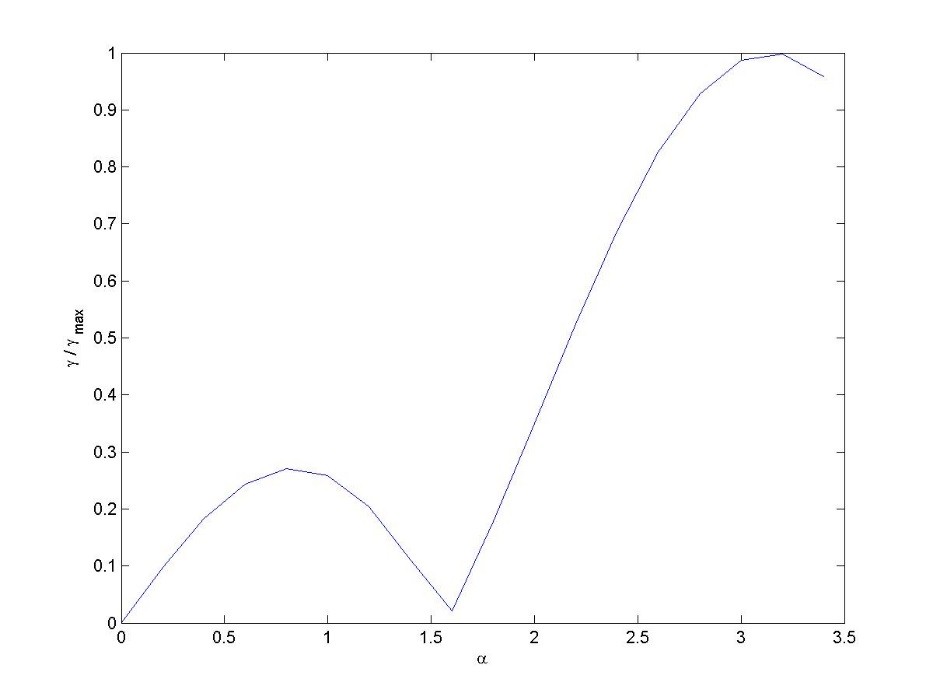}
\caption{Variation in normalized SBS instability ${\gamma}/{\gamma_{max}}$ with $\alpha$ (radian). The angle between incident light wave and scattered light wave.}
\label{FigPA}
\end{figure}
The instability growth rate was investigated for two incident wave frequencies. $\omega_{0}=70$GHz and $\omega_{0}=110$ GHz .(Since Electron Bernstein Wave (EBW) heating is used in this range of frequency for MAST tokamak~\cite{Jose}) These two frequencies are placed in dominant range for stimulated Brillouin scattering ($\omega_{pe}\simeq 60$ GHz $<\omega_{0}<2\omega_{pe}\simeq$ $120$ GHz). Time evolution shows absorption of pump wave in plasma. As a result, damping in instability rate is perceptible.

\begin{figure}[H]
\centering
\includegraphics[width=0.75\textwidth]{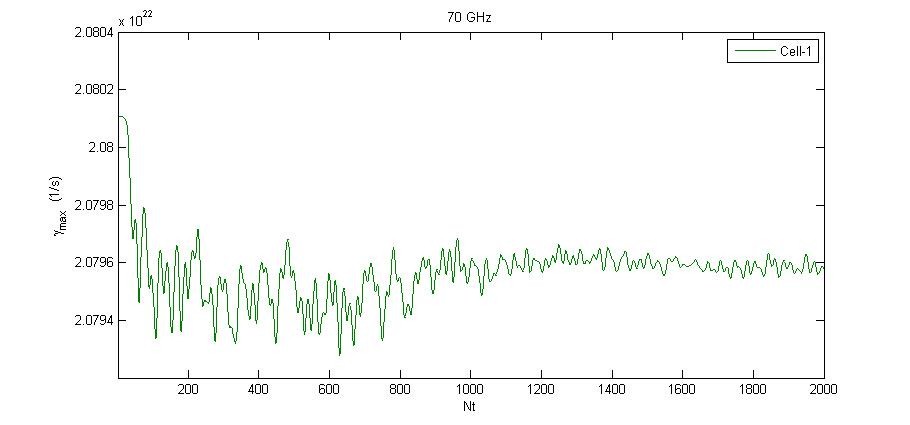}
\caption{Variation in maximum SBS instability growth rate $\gamma_{max}$ with pump frequency  $\omega_0=70$ GHz and $\omega_{pe} t=2000$ for cell $1$.}
\label{FigPA}
\end{figure} 
\begin{figure}[H]
\centering
\includegraphics[width=0.75\textwidth]{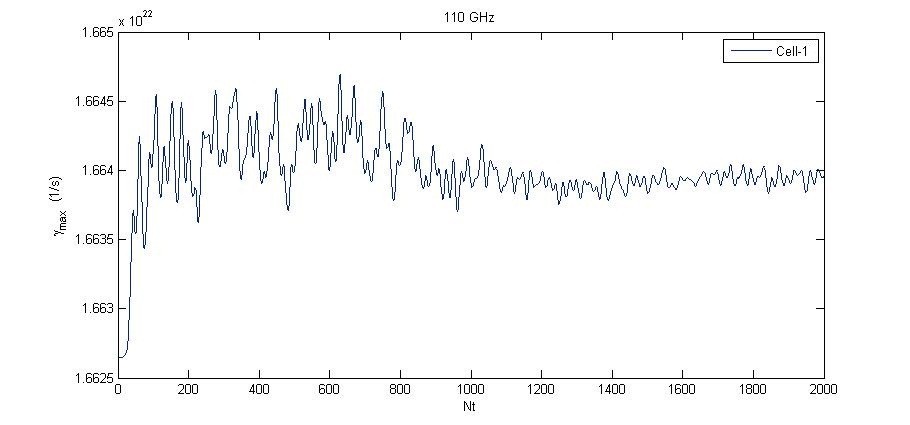}
\caption{Variation in maximum SBS instability growth rate $\gamma_{max}$ with pump frequency  $\omega_0=110$ GHz and $\omega_{pe} t=2000$ for cell $1$.}
\label{FigPA}
\end{figure} 
The maximum SBS instability growth rate is plotted by changes in frequency of incident light wave in FIG.4. The maximum instability growth rate decreases by increasing the incident light frequency for SBS clearly. Also, for frequencies more than $120$ GHz Stimulated Raman Scattering (SRS) is plotted and similarly it decreases with increasing of pump frequency~\cite{Nagh}.
\begin{figure}[H]
\centering
\includegraphics[width=0.75\textwidth]{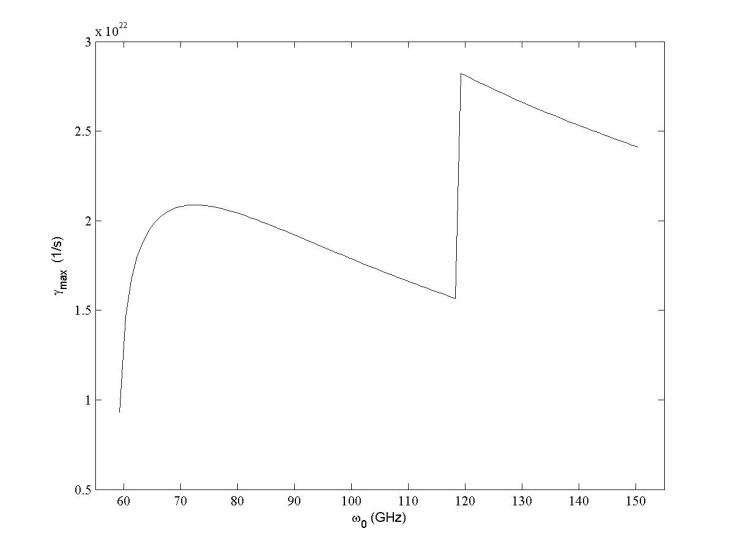}
\caption{Variation in maximum SBS instability growth rate $\gamma_{max}$ with pump frequency  $\omega_0$ and $\omega_{pe} t=2000$ for cell $50$.}
\label{FigPA}
\end{figure} 
Since the incident light wave $\omega_{0}$  is considered as O-Mode, for $\omega_{0}<\omega_{pe}=60$ GHz we are in cut-off area and our wave cannot penetrate into the plasma ($n \approx n_{cr}$). Furthermore, for frequencies more than $120$ GHz, SBS is not dominant and Stimulated Raman scattering (SRS) is more effective ($n=0.25n_{cr}$).
The maximum growth rate instability (backscattering) was investigated in $3$ arbitrary time steps ($\omega_{pe} t\equiv N_{t}=1$, $\omega_{pe} t\equiv N_{t}=20$ and $\omega_{pe} t\equiv N_{t}=2000$ ) in order to examine the effect of pump wave on $x$ steps more precisely. At $\omega_{pe} t=1$, we do not have any oscillations in the cells, since our wave has not entered into the plasma yet. By evolution of time we have oscillations in all $50$ cells which we can see it easily at $\omega_{pe} t=2000\equiv N_{t}$.
\begin{figure}[H]
\centering
\includegraphics[width=0.75\textwidth]{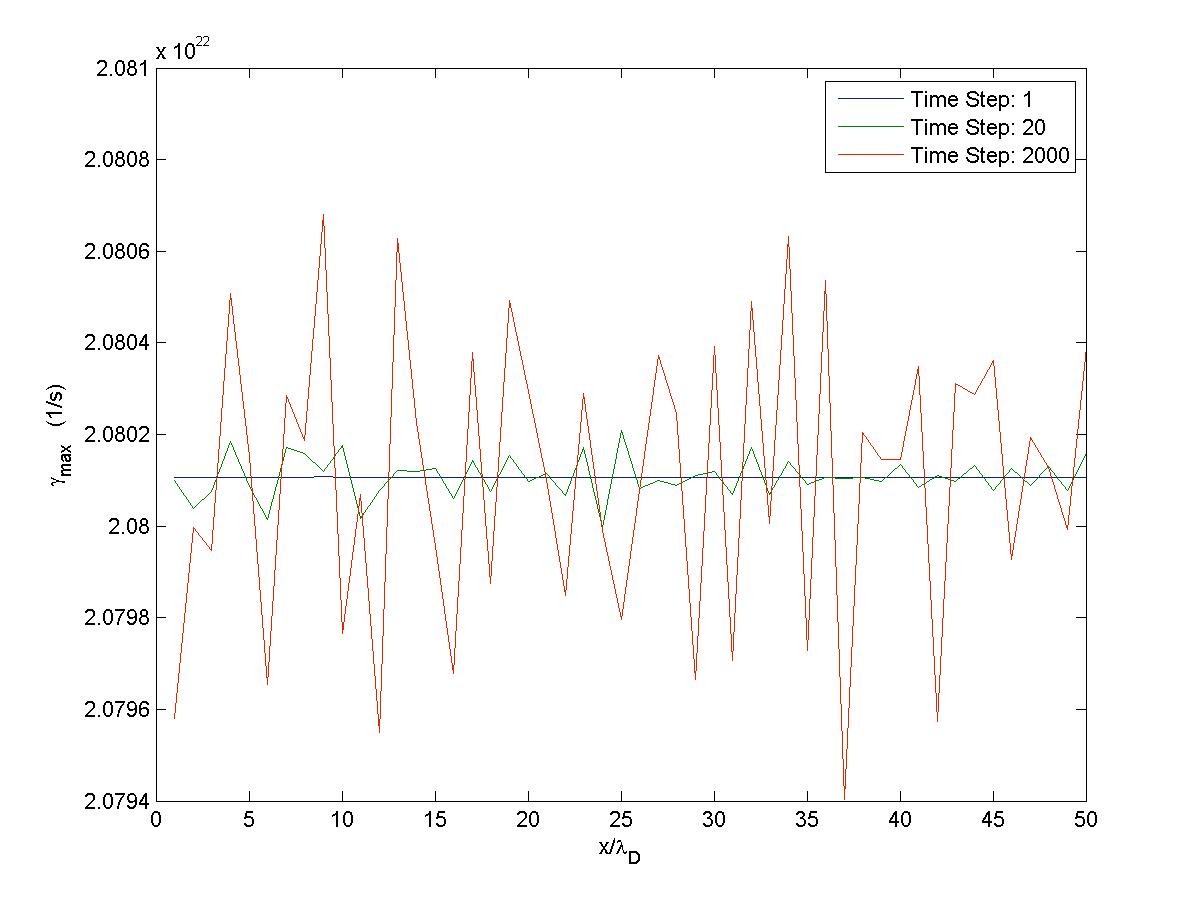}
\caption{ Maximum SBS instability growth rate $\gamma_{max}$ with $\omega_0=70$ GHz and $\omega_{pe} t=1,20,2000$ .}
\label{FigPA}
\end{figure}
Also, time evolution of maximum instability growth rate is plotted for $\omega_0=70$ GHz in $2$ different cells ($1$ and $50$).  Damping is clear by time evolution, since the pump wave's energy is absorbed in plasma (plasma heating).
\begin{figure}[H]
\centering
\includegraphics[width=0.75\textwidth]{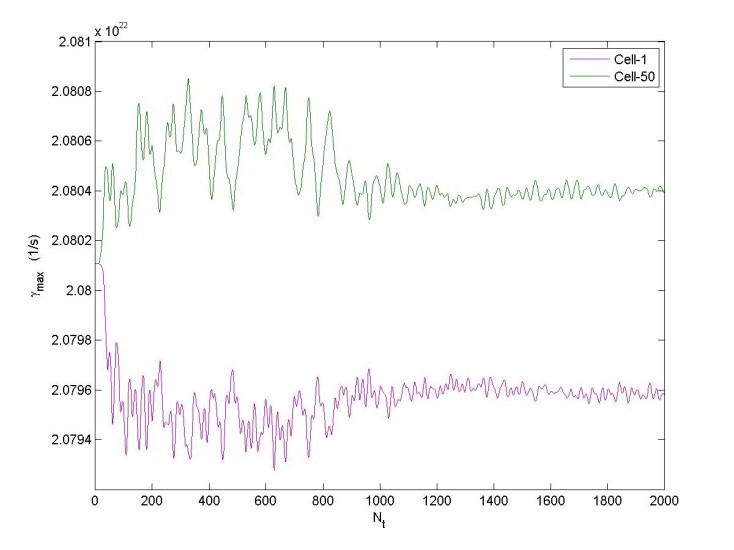}
\caption{Time evolution of maximum instability growth rate $\gamma_{max}$ is plotted with $\omega_0=70$ GHz and $\omega_{pe} t\equiv N_{t}=2000$  for cells $1$ and $50$.}
\label{FigPA}
\end{figure}
 Since Landau damping causes IAW to damp and increase the heating, dependence of Landau damping on ion temperature is investigated as well. The equation for this damping is:
\begin{equation}
\omega_{i}=-\frac{\arrowvert \omega_{r} \arrowvert}{(1+k^2 \lambda_{De}^2)^{3/2}}\sqrt{\frac{\pi}{8}}[\sqrt{\frac{m_{e}}{m_{i}}}+\big(\frac{T_{e}}{T_{i}})^{3/2} \exp \big(-\frac{T_{e}/2T_{i}}{1+k^2 \lambda_{De}^2}-\frac{3}{2})
\end{equation}
\begin{equation}
\omega_{r}^2=\frac{k^2 c_{s}^2}{1+k^2\lambda_{De}^2}+3k^2\frac{K_{B}T_{i}}{m_{i}}
\end{equation}
Which $\omega_r$ and $\omega_i$ are the real and imaginary parts of the frequency, respectively.~\cite{Bell}\\
 As we expected, when $T_i$  approaches  $T_e$ , this value is enhanced. While for $T_e\gg T_i$ damping is decreased.
\begin{figure}[H]
\centering
\includegraphics[width=0.75\textwidth]{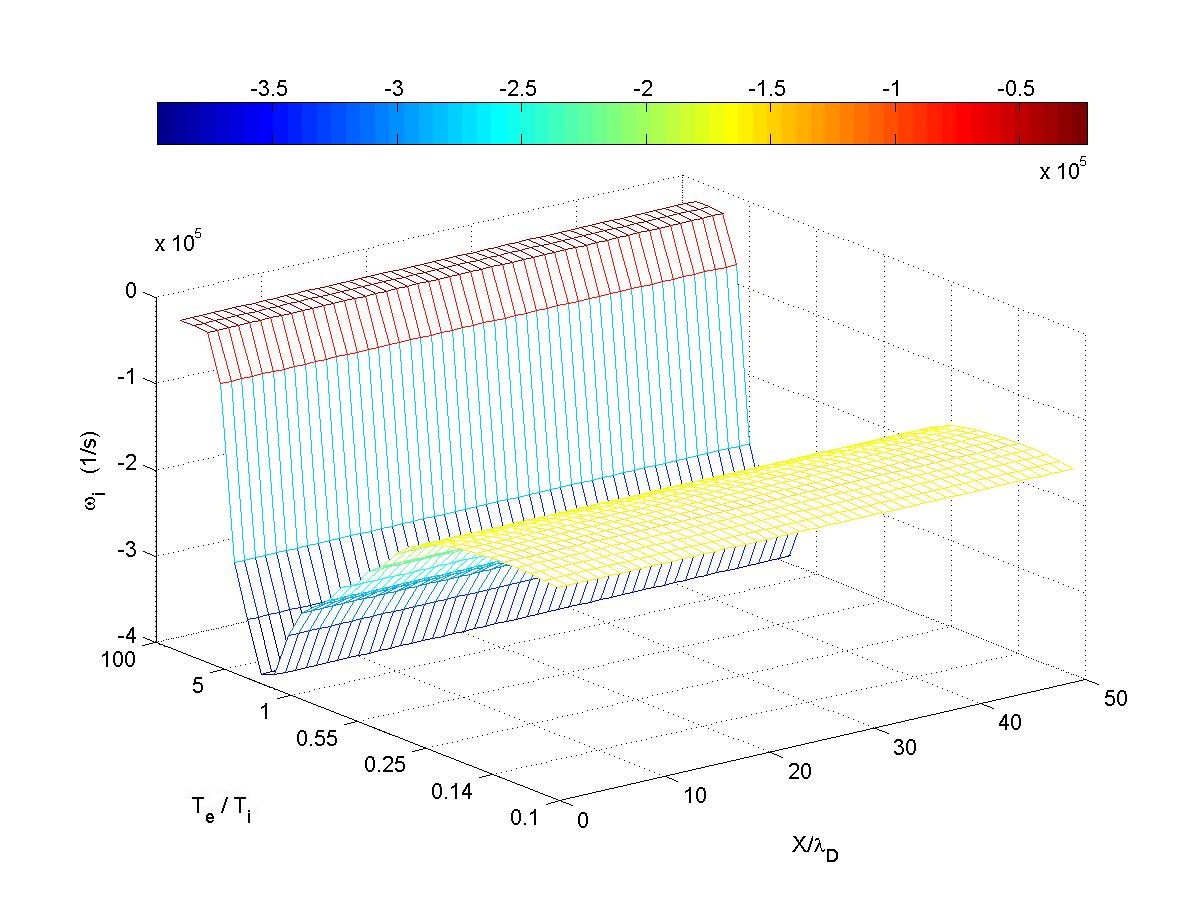}
\caption{Variation in Landau damping for IAW $\omega_i$ with electron-to-ion temperature ratio ${T_e}/{T_i}$  for all cells and $\omega_{pe} t=2000$ .}
\label{FigPA}
\end{figure}
\section{IV.Discussion:}
Instabilities always occur in plasma and SBS is one of them. Since our goal is plasma heating and in SBS a big portion of incident wave's energy transfers to scattered EM wave hence, \textquotedblleft this instability is a significant concern for laser fusion applications" ~\cite{Kru}. The instability rate depends on distribution function and this function is different in all cells and velocities. Increasing the frequency of pump wave causes the instability growth rate to decrease. Furthermore, after injection of the pump wave, Landau damping causes IAW to damp. As a result, conditions that increase it like electron-to-ion temperature ratio, is important to investigate. Also, the growth rate of Brillouin backward instability is investigated in unmagnetized plasma. Investigation of it in magnetized plasma \textquotedblleft shows a decrease due to the presence of external magnetic field~\cite{Pak}".
\section{V.Reference:}

\end{document}